\documentclass{emulateapj}
\usepackage{epsfig}
\usepackage[dvips]{color}

\newcommand{\abs}[1]{\left\vert #1\right\vert}
\def\alt{\raise0.3ex\hbox{$\;<$\kern-0.75em\raise-1.1ex\hbox{$\sim\;$}}}
\def\agt{\raise0.3ex\hbox{$\;>$\kern-0.75em\raise-1.1ex\hbox{$\sim\;$}}}

\definecolor{Black}{named}{Black}
\definecolor{Red}{named}{Red}

\newcommand{\bw}{\begin{widetext}}
\newcommand{\ew}{\end{widetext}}

\begin{document}

\title{Secondary radiation  from the Pamela/ATIC excess and relevance for Fermi}
\author{
E.~Borriello$^{1}$, A.~Cuoco$^2$ and G.~Miele$^{1}$}

\affil{
    $^1$Dipartimento di Scienze Fisiche, Universit\`a di Napoli ``Federico II'' \& INFN Sezione di Napoli, \\ Complesso
    Universitario di Monte S. Angelo, Via Cinthia 80126, Napoli, Italy \\
    $^2$Department of Physics and Astronomy, University of Aarhus, Ny Munkegade, Bygn. 1520 8000, Aarhus, Denmark \\
}


\begin{abstract}
The excess of electrons/positrons observed by the Pamela and ATIC
experiments gives rise to a noticeable amount of synchrotron and
Inverse Compton Scattering (ICS) radiation when the $e^+e^-$
interact with the Galactic Magnetic Field, and the InterStellar
Radiation Field (ISRF). In particular, the ICS signal produced
within the WIMP annihilation interpretation of the Pamela/ATIC
excess shows already some tension with the EGRET data. On the other
hand, 1 yr of Fermi data taking will be enough to rule out or
confirm this scenario with a high confidence level. The ICS
radiation produces a peculiar and clean ``ICS Haze'' feature, as
well, which can be used to discriminate between the astrophysical
and Dark Matter scenarios. This ICS signature is very prominent even
several degrees away from the galactic center, and it is thus a very
robust prediction with respect to the choice of the DM profile and
the uncertainties in the ISRF.
\end{abstract}


\maketitle

The Pamela and ATIC results have recently raised a great interest in
the scientific community due to the possibility that the observed
$e^+e^-$ excesses could be a signature of the, so-far elusive,
particle associated to Dark Matter. The raise in the positron
fraction above 10 GeV until $\sim$100 GeV seen by Pamela
\citep{Adriani:2008zr} and the excess of the sum of $e^+$ and $e^-$
between $\sim$100 GeV and $\sim$700 GeV seen by ATIC \cite{:2008zzr}
can be hardly explained in a standard Cosmic Ray production scenario
and, instead, seem to point to a new source of $e^+$ and $e^-$.
Hints of this anomaly were reported also by different experiments
like HEAT \cite{Barwick:1997ig}, AMS-01
\citep{Aguilar:2007yf,Alcaraz:2000bf} and PPB-BETS
\cite{Torii:2008xu}. In addition, HESS has recently presented a
measurement of the electron spectrum in the range $0.6<E<5$ TeV
\cite{Collaboration:2008aaa}. This anomaly can have a standard
astrophysical interpretation (\citealp{atoian},
\citealp{ZhangCheng}, \citealp{profumo},\citealp{Yuksel},
\citealp{hpb}) or an exotic one involving decaying (\citealp{liu},
\citealp{Hisano}, \citealp{Yin}, \citealp{Chen}, \citealp{Ibarra},
\citealp{Hamaguchi}) or the annihilation of DM particles
(\citealp{Hisano2008}, \citealp{Mardon}, \citealp{Zurek},
\citealp{Cholisb}, \citealp{Bergstrom}, \citealp{ArkaniHamed},
\citealp{Meade:2009rb},~\citealp{Ishiwata},~\citealp{Hu:2009bc},
\citealp{Nomura}, \citealp{Hall},
\citealp{Barger},~\citealp{deBoer},~\citealp{Cholisc},
\citealp{Fox}). The latter description, in particular, seems to
favor a DM particle in the TeV range and with a thermally averaged
annihilation cross section $\left\langle \sigma_{\mathrm{A}}v\right
\rangle \sim 10^{-23}$ cm$^3$s$^{-1}$. However, this scenario faces
several difficulties. A first problem is that, differently from the
positron ratio, no excess is observed by Pamela in the antiproton
over proton ratio \cite{Adriani:2008zq}. This means that DM
decay/annihilation into hadronic channels is mainly forbidden or at
least strongly suppressed \cite{Cirelli:2008,Donato:2008jk}, and
hence one has to resort to models in which only the leptonic
channels are allowed. The second problem is that the annihilation
rate required to explain the anomaly is about three orders of
magnitude above the natural expectation of $\left<
\sigma_{\mathrm{A}}v\right>\sim 3\times10^{-26}$ cm$^3$s$^{-1}$ for
a DM thermal relic which accounts for the cosmological DM abundance.
This requires either the introduction of large annihilation boost
factors from the presence of galactic substructure, or some
enhancing annihilation mechanism like the Sommerfeld
process~\cite{lattanzi,ibe}.

\begin{figure*}[t]
 \vspace{-1.0pc}
\begin{center}
\begin{tabular}{cc}
\vspace{-0.5pc}\includegraphics[width=0.95\columnwidth,angle=0]{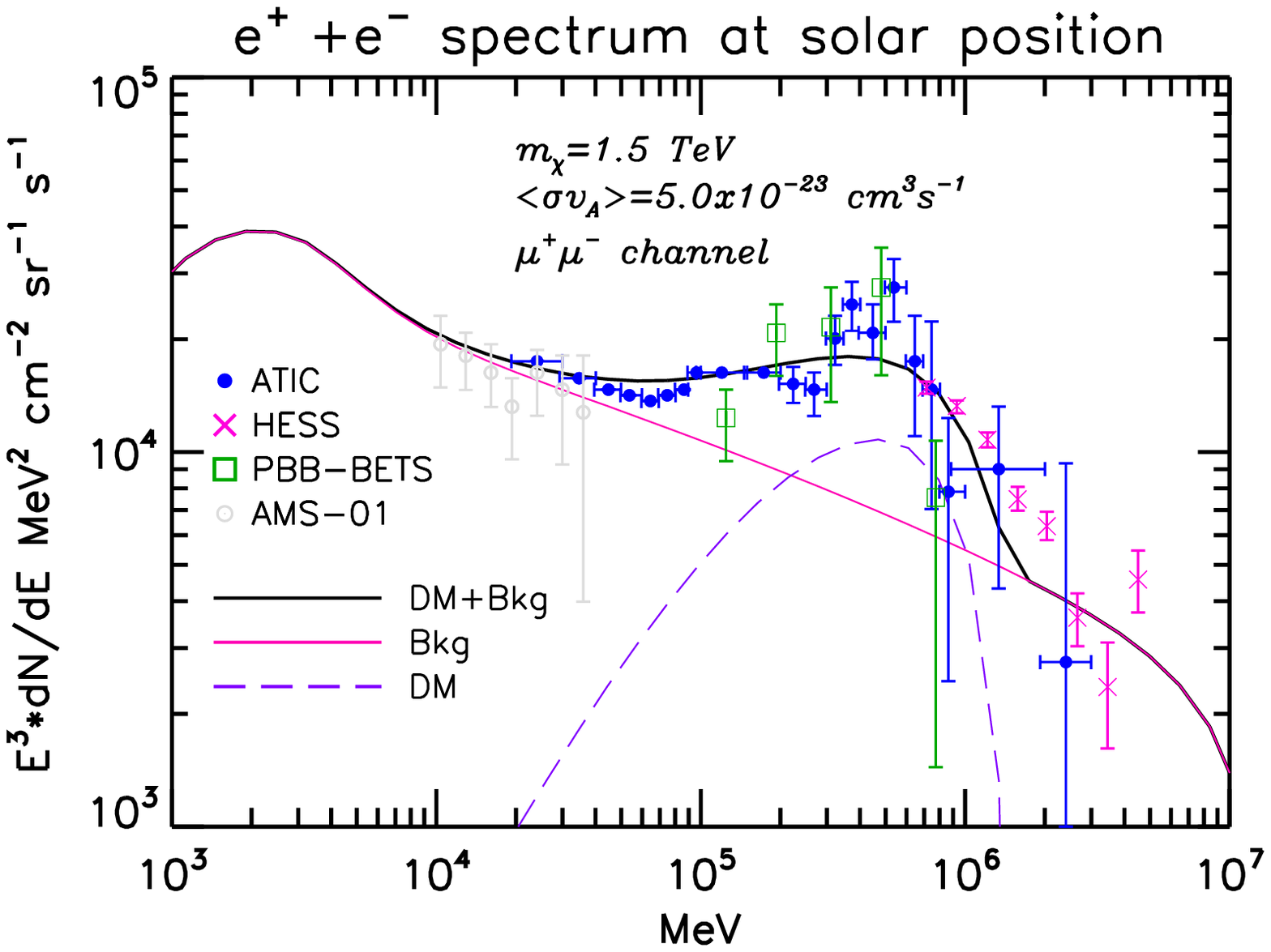}
&
\vspace{-0.5pc}\includegraphics[width=0.95\columnwidth,angle=0]{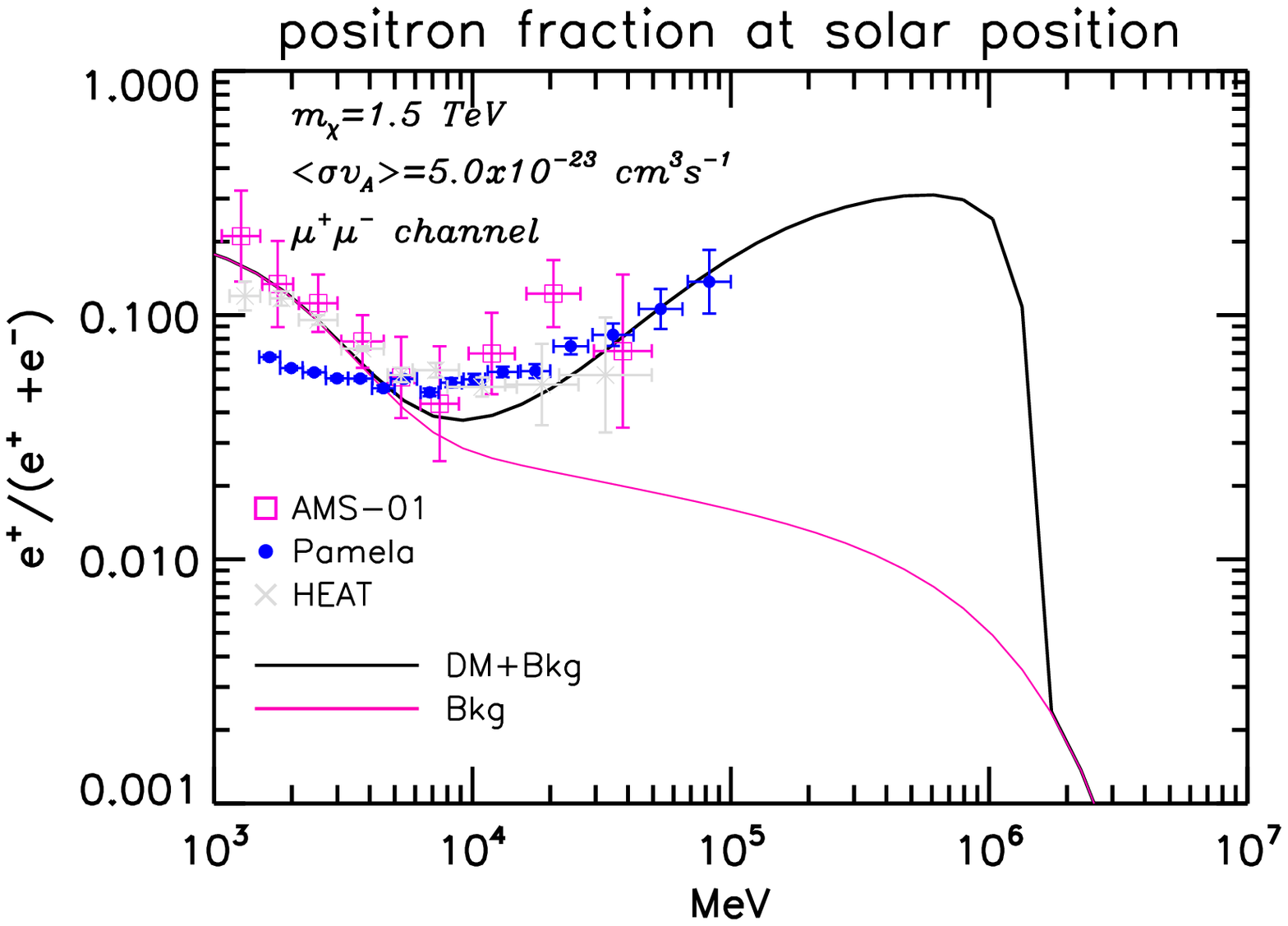}\\
\vspace{-0.5pc}\includegraphics[width=0.95\columnwidth,angle=0]{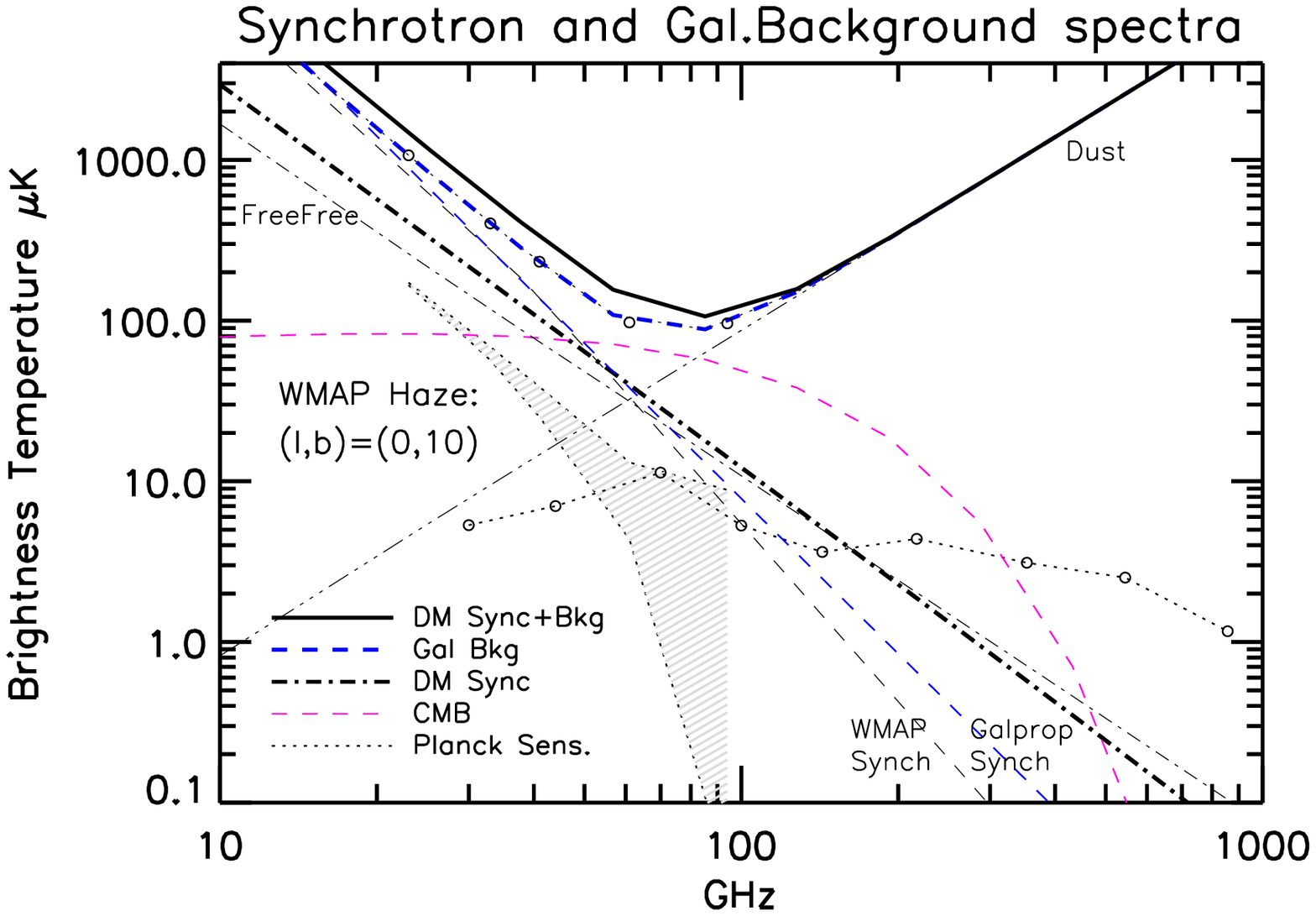}
&
\vspace{-0.5pc}\includegraphics[width=0.95\columnwidth,angle=0]{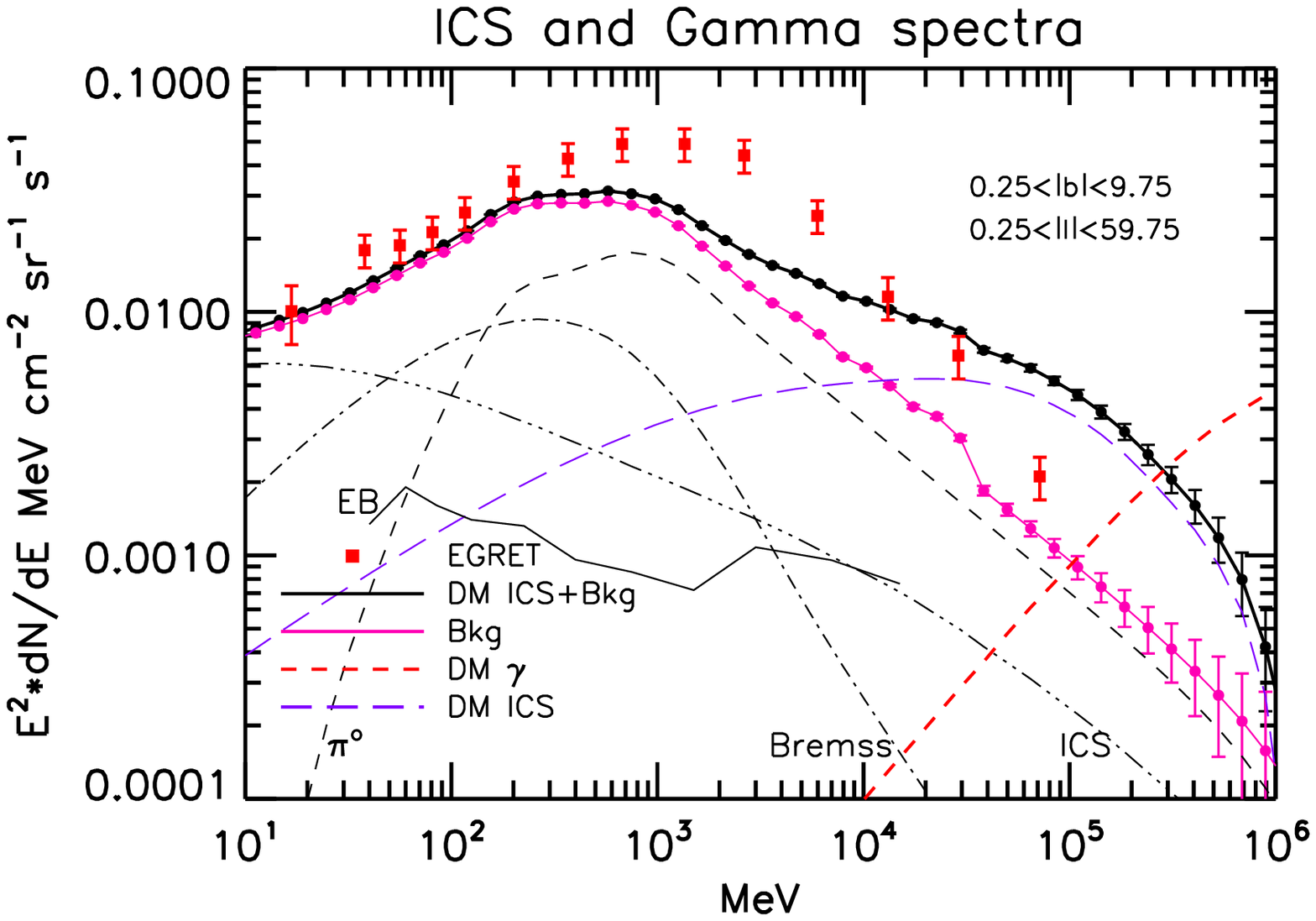} \\
\end{tabular}
\vspace{-0.5pc}
\end{center}
\hspace{-1.5pc} \caption{\scriptsize{The upper panels show the
positron fraction and the total $e^+e^-$ spectrum for the CR
background and the DM annihilation signal compared with the Pamela
and ATIC data. A compilation of previous data
(HEAT~\cite{Barwick:1997ig} and AMS-01~\cite{Aguilar:2007yf} for the
positron fraction and PPB-BETS~\cite{Torii:2008xu},
AMS-01~\cite{Alcaraz:2000bf} and HESS~\cite{Collaboration:2008aaa}
for the $e^+e^-$ spectrum) is also shown. The lower right panel
reports the gamma spectrum for the CR background and the ICS signal
from DM electrons in the Halo together with the EGRET measurements
and the errors expected after a 1 yr survey by Fermi. The red dashed
curve shows the spectrum of gamma-rays produced directly through the
annihilation into $\mu^+\mu^-$. The decomposition of the CR
background into the IC, bremsstrahlung, pion decay and extragalactic
components is reported as well. The lower left panel shows the DM
synchrotron emission, in units of brightness temperature, $10^\circ$
away from the GC compared with the galactic backgrounds as measured
by WMAP~\cite{Gold:2008kp} and the r.m.s. fluctuations of the CMB.
The open points indicate the 9 Planck frequencies while the dotted
line shows the expected Planck sensitivity for a 14 months
survey~\cite{:2006uk}. The second set of open points indicates the
WMAP frequencies. For comparison it is shown the signal from the
WMAP Haze $10^\circ$ degrees away from the GC as derived in
\cite{Dobler:2007wv}. Furthermore we report the decomposition of the
Galactic backgrounds into the dust, free-free and synchrotron
components together with the synchrotron background derived with
Galprop. A model with a WIMP of $m_\chi=1.5$ TeV which annihilates
only into $\mu^+\mu^-$ with a rate $\left\langle
\sigma_{\mathrm{A}}v\right\rangle\sim 5\times10^{-23}$cm$^3$s$^{-1}$
is considered for all the plots. The propagation parameters are
specified in the text.} \label{fig:1}}
\end{figure*}

The fact that hadronic channels have to be suppressed to explain the
Pamela/ATIC anomaly implies that only few (energetic) photons are
produced either if the annihilation takes place through the
$\mu^+\mu^-$ or $\tau^+\tau^-$ channels or in the case of the
$e^+e^-$ channel through the presence of Final State Radiation. With
the limited contribution of gamma rays accompanying the annihilation
process, the constraints from gamma observations become thus quite
weak. Anyway, even though only $e^+e^-$ were produced in the DM
annihilation process, these leptons, once in the galactic
environment, would interact with the Galactic Magnetic Field (GMF)
and the Interstellar Radiation Field (ISRF). Thus they would lose
energy producing synchrotron radiation in the radio band and Inverse
Compton Scattering (ICS) Radiation in the gamma band. This secondary
radiation thus represents a complementary observable to constrain
the DM signal
(\citealp{Bergstrom:2008ag},~\citealp{Ishiwata:2008qy},
\citealp{Cholis:2008wq},~\citealp{Nardi:2008ix},~\citealp{Zhang:2008tb},
 \citealp{Bertone:2008xr},
\citealp{Borriello:2008gy}). In the following we will focus on the
synchrotron and ICS signals which are expected in the galactic
halo. With respect to focussing on the Galactic Center (GC) this
approach provides much more robust predictions due to the weaker
dependence on the choice of the DM profile and thanks to the
smaller uncertainties on ISRF and GMF. The relevance of ICS signal
in relation to Pamela has been, indeed, discussed in recent papers
\cite{Zhang:2008tb,Cholis:2008wq} which show the presence of some
tension with the EGRET data as well. In the following, we will
stress how the situation is expected to change with the new data
from Fermi and, further, we will investigate the peculiar spatial
distribution which the DM signal is expected to produce.

We use for the calculations a slightly modified version of Galprop
v50.1p \cite{Strong:1998pw,Moskalenko:1997gh}, which  solves
numerically the electron diffusion-loss equation and produces the
ICS and synchrotron maps. The code also provides maps of the CR
gamma diffuse emission using available data on the CR abundances
and the distribution of galactic gas. For our calculations we
employ a diffusion coefficient $D=D_0 (E/E_0)^{-\alpha}$ with $D_0
= 5\times10^{28}$ cm$^2$s$^{-1}$, $E_0=3$ GeV and $\alpha=0.33$,
corresponding to a Kolmogorov spectrum of turbulence. The
transport equation is solved in a cylinder of half-height $z=\pm
4$ kpc and radius $R=20$ kpc, while the GMF used to derive the
synchrotron radiation is modeled as $\left<B^{2}\right>^{1/2}= B_0
\exp(-r/r_B-\abs{z}/z_B)$ with $B_0=11 \mu$G $r_B=10$ kpc and
$z_B=2$ kpc. It is worth reminding, however, that electrons have
typically a quite short propagation length (in terms of the
galactic size) corresponding to a path of $\mathcal{O}$(1 kpc)
\cite{Delahaye:2007fr} before losing a significant percentage of
their energy. Thus the final spectrum and distribution of
electrons keep only a weak dependence on the chosen propagation
parameters. The GMF, on the other hand, is still affected by large
uncertainties especially in the inner kpc's of the galaxy (see
\cite{Han:2009ts} and reference therein for a recent review). The
synchrotron radiation, which is quite dependent on the GMF,
shares, thus, a similar uncertainty on the normalization. The
InterStellar Radiation Field, which is the photon target that
determines the ICS signal, is, instead, better known and the
derived ICS signal is thus a more robust prediction than the
synchrotron signal. The ISRF implemented in Galprop is described
in details in \cite{Porter:2005qx}. Finally, for the DM profile we
choose a very conservative isothermal cored one, namely
$\rho(r)=\rho_0\left(r_c^2+r_\odot^2\right)/\left(r_c^2+r^2\right)$,
with a DM density $\rho_0=0.4$ GeV/cm$^3$ at the solar position
$r_\odot=8.5$ kpc. We fix $r_c=2.8$ kpc for the the core size.
However, this particular choice is not crucial since we are going
to calculate the signal not in GC, but the one coming from the
Halo where the uncertainties on the details of DM profile are less
relevant.

\begin{figure}[t]
\vspace{0pc}
\begin{center}
\begin{tabular}{c}
\includegraphics[width=.50\columnwidth,
angle=90]{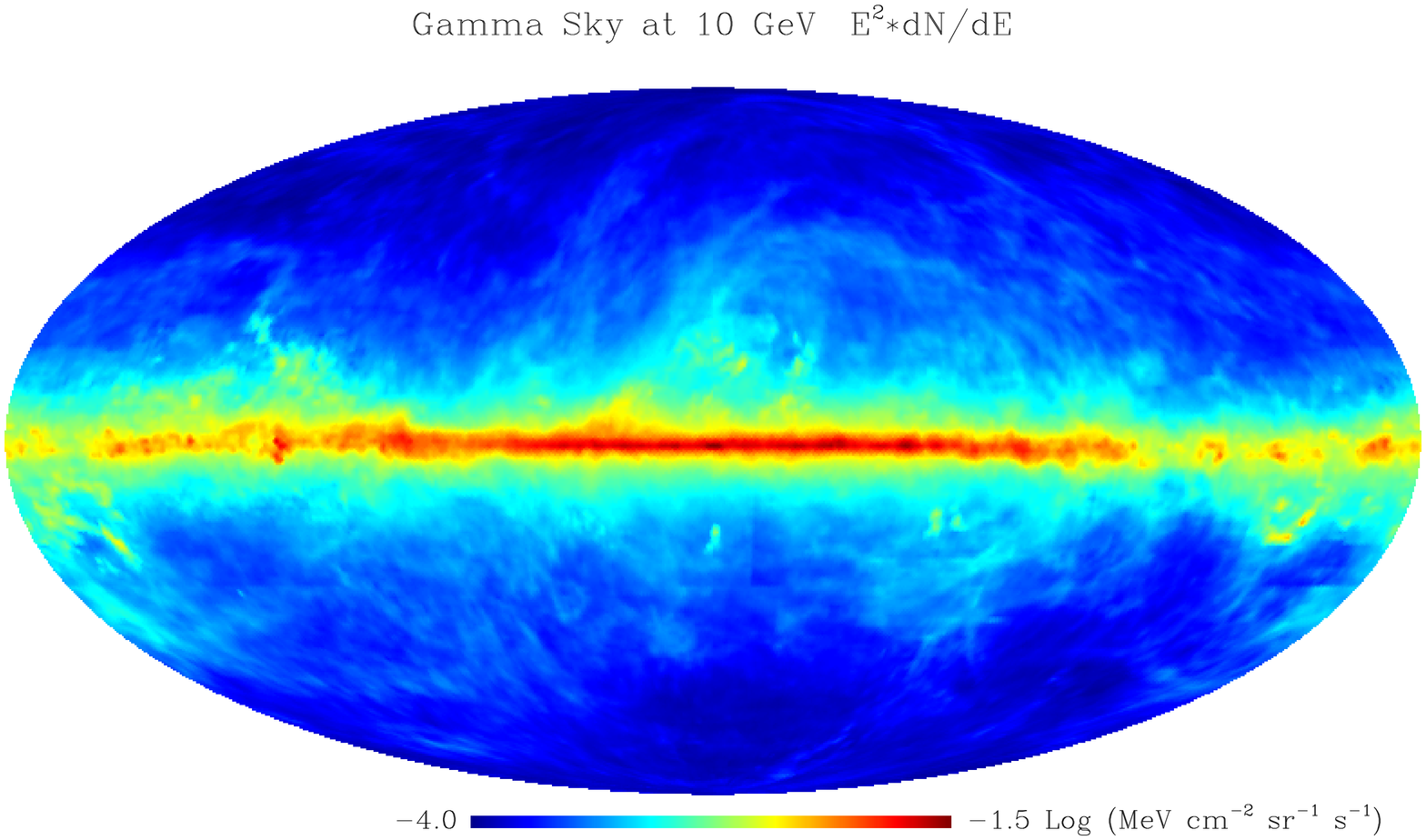}
\\\includegraphics[width=.50\columnwidth,
    angle=90]{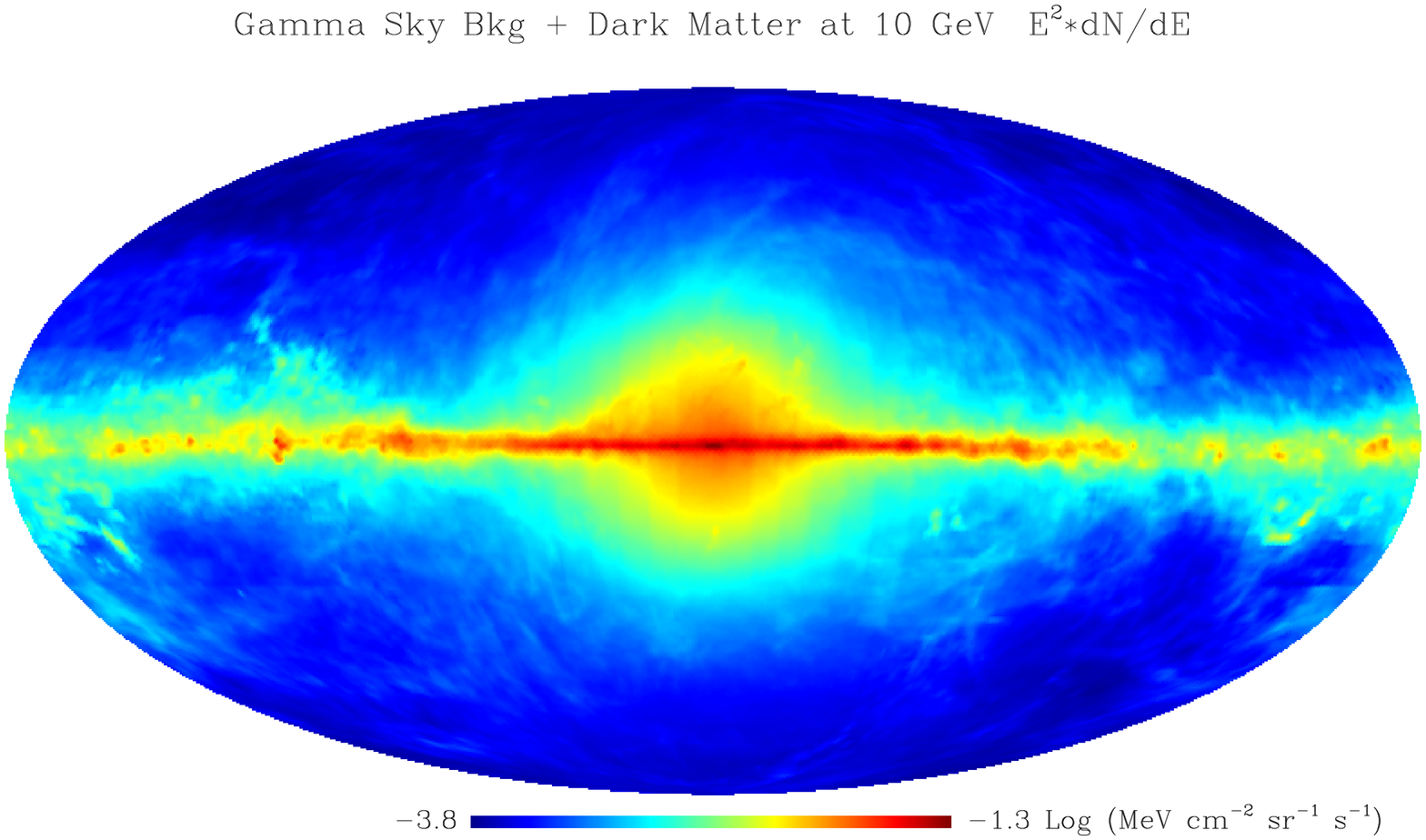} \\
\includegraphics[width=.50\columnwidth,
    angle=90]{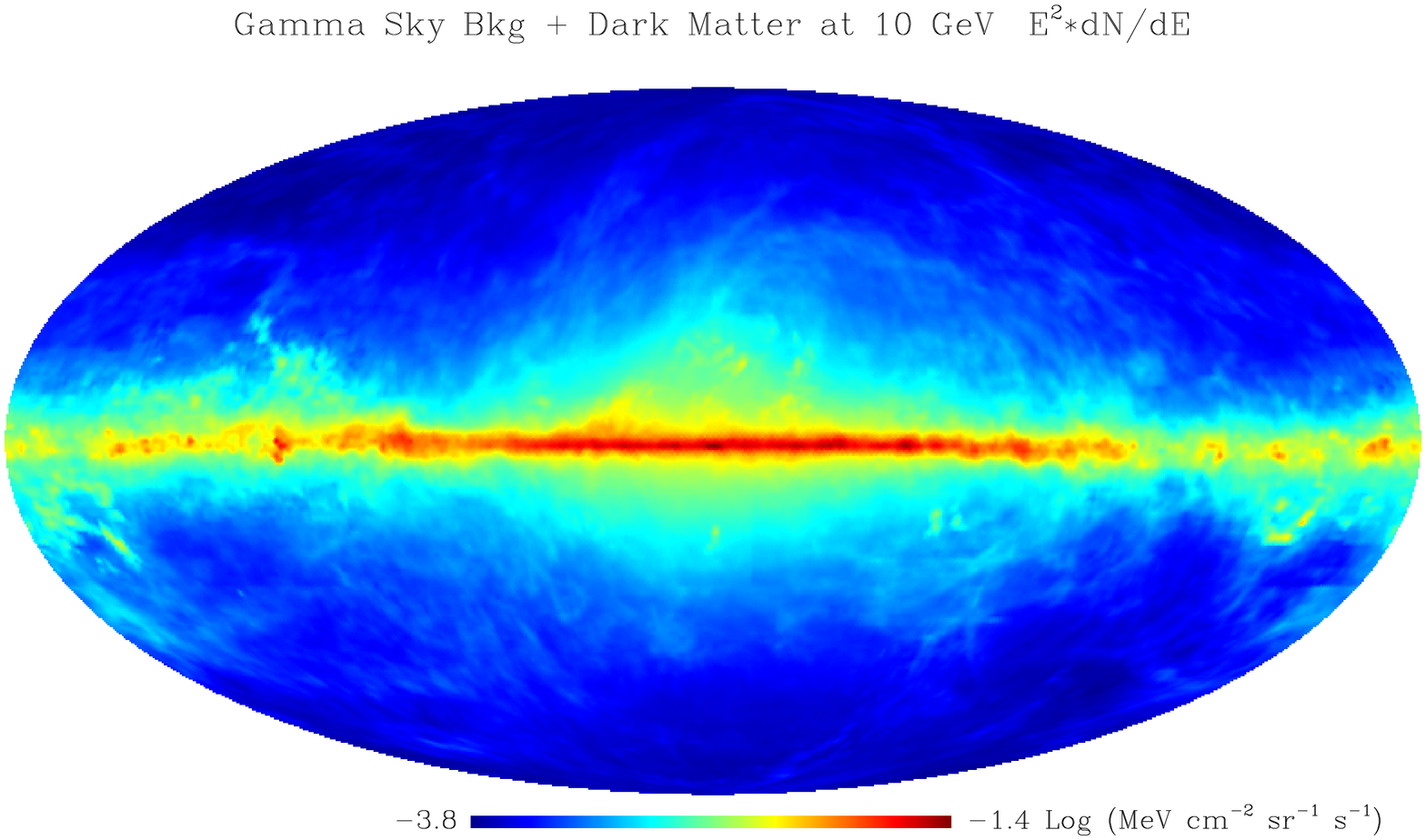}
\end{tabular}
\end{center}
\vspace{-1pc}
        \caption{Sky map (in healpix format \cite{Gorski:2004by}) of galactic gamma backgrounds at the energy of
        10 GeV (top).  The same with the inclusion of the DM annihilation contribution (center)
        or Decaying  DM (bottom). \label{fig:Skymaps}}
\vspace{0pc}
\end{figure}

We choose to study a single benchmark model with a WIMP of
$m_\chi=1.5$ TeV annihilating in the $\mu^+\mu^-$ channel only with
a rate $\left\langle \sigma_{\mathrm{A}}v\right\rangle\sim
5\times10^{-23}$cm$^3$s$^{-1}$. The resulting electron/positron
injection spectrum $dN_e/dE$ has approximately a constant behavior
in energy with a cutoff at the mass of the WIMP. We calculate
$dN_e/dE$ with DarkSUSY~\cite{Gondolo:2004sc}, which, in turn, uses
a tabulation of the spectrum of the decay products derived with
Pythia~\cite{Sjostrand:2007gs}. The electron source term for Galprop
is then given by $Q(r,E)=\rho^2\left\langle
\sigma_{\mathrm{A}}v\right\rangle/2m_\chi^2\times dN_e/dE$. This
model provides a reasonable good match with the Pamela and ATIC
data. It is certainly possible to achieve a better fit with a mixing
of the various leptonic channels, or with particular alternative
annihilation mechanism or, further, with a fine tuning of the
propagation parameters \footnote{\footnotesize{During the review
procedure of our paper the Fermi collaboration has reported a
measurement of the $e^+e^-$ flux in the same energy range of ATIC
\cite{Collaboration:2009zk}. The spectrum measured by Fermi confirms
an excess with respect to the conventional cosmic ray model although
the excess is less prominent and smoother than the one reported by
ATIC. For this broad smooth excess a better fit can be achieved
through an annihilation into $\tau^+\tau^-$ instead of $\mu^+\mu^-$.
Using the $\tau^+\tau^-$ channel as benchmark model, however,
produces just minor changes in the results derived in the
following.}}. However, since the aim of our paper is to focus on the
secondary radiation, the final results would be only weakly affected
by the above details on the WIMPs annihilation process. The results
are illustrated in Fig.\ref{fig:1}. The upper panels show the
comparison of the model with the Pamela and ATIC data and with a
compilation of previous data showing that, indeed, the agreement is
good. The secondary radiation results are shown in the lower panels.
The right one shows the expected difference between the CR gamma
background and the ICS produced by the population of DM electrons
distributed in the galactic halo together with the EGRET
measurements (as taken from~\cite{Strong:2004de}). Furthermore the
decomposition of the CR background into the IC, bremsstrahlung, pion
decay and extragalactic components is also shown. The extragalactic
component~\cite{Sreekumar:1997un} is from the reanalysis of the
EGRET data from \cite{Strong:2004ry}. The small error bars are a
forecast for $T$=1 yr of data taking by Fermi assuming the effective
area as function of energy as taken from \cite{Atwood:2009ez}
(roughly $A_{eff}$= 8000 cm$^2$ above $\sim$1 GeV) a field of view
of 2.4 sr and no CR contamination hence $N_\gamma=T\times {\rm fov}
\times f_\Delta \times\int_{\Delta E}A_{eff}(E) dN_\gamma/dE(E) dE$.
$dN_\gamma/dE(E)$ is the gamma ray flux while $f_\Delta$ is the
fraction of area of the sky where the signal is integrated.  The
Poisson error is then $\propto 1/\sqrt{N_\gamma}$. Finally the
errors are shown for a logarithmic binning of the energy.

It is worth noticing that the errors expected for one year from
Fermi survey are tiny enough to detect the excess with an high
degree of confidence. Even more importantly, this excess comes
from the halo region, placed several degrees away from the GC and
thus in a region were the uncertainties on the DM profile are
expected to be much smaller. Also the uncertainty on the ISRF,
which seems anyway not critical \cite{Porter:2005qx}, naturally
decreases moving away from the GC. A possible problem is, in
principle, the fact that the DM excess can be mistaken with a not
well understood CR gamma background. Indeed, the situation is
similar to the EGRET GeV excess \cite{Hunger:1997we} which, in
principle can be explained either with an ``optimized'' CR model
\cite{Strong:2004de} or with a DM contribution
\cite{deBoer:2005tm}\footnote{Note, anyway, that preliminary
results from the Fermi collaboration seem not to confirm the GeV
excess. See e.g. the talk presented on behalf of the Fermi
collaboration at the January 2009 meeting of the AAS. }. In this
case, however, the IC excess produced by Pamela/ATIC is more
properly a ``10-100 GeV excess''. Moreover, it generally exceeds
already the EGRET data, although by an amount which is still in
principle within the EGRET systematics. A more crucial difference
is however the spatial distribution. While the GeV excess is
almost isotropic in the sky, the ICS excess has the shape of a
circular Haze reflecting the DM distribution in the Halo. This
difference, indeed, is quite striking, as can be seen clearly in
Fig.\ref{fig:Skymaps}. The CR background instead is expected to
lie mostly along the galactic plane where the astrophysical
sources are located.

The lower left panel shows the DM synchrotron emission in units of
brightness temperature ($T\propto \nu^{-2}F_\nu$) $10^\circ$ away
from the GC compared with the galactic backgrounds. We use the WMAP
background  maps (CMB subtracted) and their decomposition into
synchrotron, free-free and dust \cite{Gold:2008kp}\footnote{Data are
available at the Lambda web site: http://lambda.gsfc.nasa.gov/}. For
illustration the frequency spectra in the plot are extrapolated also
outside the WMAP frequency coverage. We also show for comparison the
background synchrotron emission calculated with Galprop which,
indeed, exhibits a close match with the WMAP synchrotron spectrum in
the 20-100 GHz range. It has to be noticed that the synchrotron
galactic CR emission dominates the background only up to a frequency
of $\sim 60$ GHz, then there is a small frequency window which is
dominated by free-free (thermal bremsstrahlung) emission, while
above $\sim$100 GHz the background is dominated by dust emission.
The fluctuations of the CMB dominates around $\sim$100 GHz depending
on the galactic latitude. The high quality data from WMAP, however,
allow to efficiently clean this further ``background''. The DM
synchrotron radiation would exhibit in principle a peak with respect
to the synchrotron background around a frequency $\sim 10^5$ GHz( as
shown in \cite{Zhang:2008tb}), where, however, the dust background
is dominating by many orders of magnitude. Restricting the analysis
in the more interesting frequency range $< 1000$ GHz, the DM signal
has an almost power law behavior with a slope slightly harder than
the background, while the spatial distribution has a circular shape.
These characteristics indeed correspond to what is found in the WMAP
Haze \cite{Dobler:2007wv,Hooper:2007kb,Cumberbatch:2009ji} whose
signal we also report in the plot for comparison. Notice, however,
that the Haze feature has still to be firmly established and that at
the moment it is very much dependent on the method employed to
separate the foregrounds \cite{Gold:2008kp}. Interestingly, we find
that, for the GMF model employed, the DM signal exceeds the Haze for
a factor of $\sim3$ similarly to the IC case. The theoretical
signal, on the other hand is affected by the uncertainties on the
GMF and it is difficult to normalize reliably. Moreover, further
uncertainties come from the systematics involved in the separations
of the measured signal into the various components, synchrotron,
dust, free-free and DM, hence it would be difficult to asses the
real significance of this excess.

\begin{figure}[!t]
\vspace{-0pc}
\begin{center}
\includegraphics[width=.90\columnwidth, angle=0]{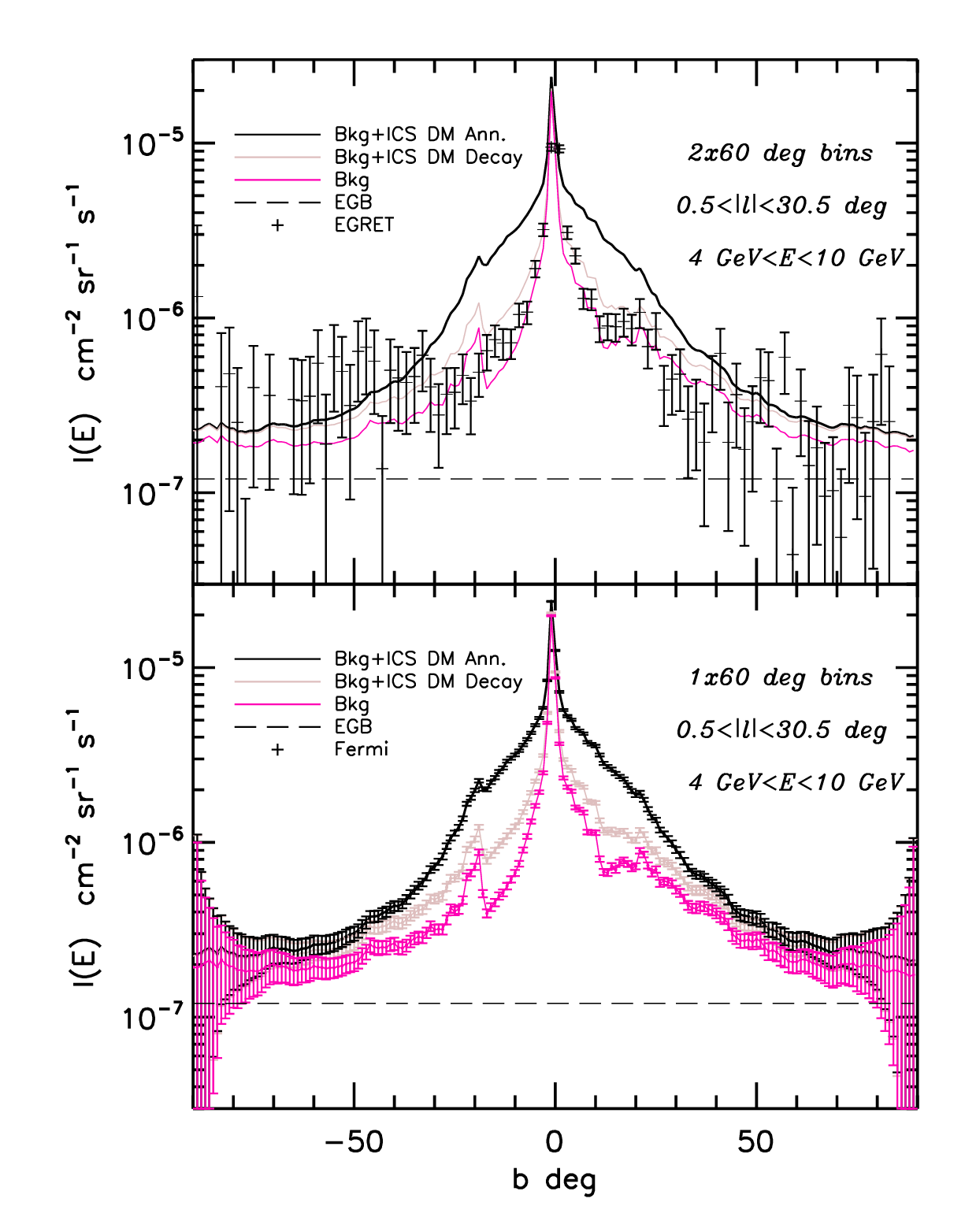}
\vspace{-1pc}
 \caption{Top panel: Background and DM (either annihilating and decaying) latitude gamma profiles
 averaged in a strip of $60^\circ$ along $l=0$ compared with the
 EGRET data. Bottom panel: same as above, but with the errors expected
 with a 1yr survey from Fermi. At high latitudes
 the error bars appear artificially to increase for the geometry of the
 $0.5^\circ<\abs{l}< 30.5^\circ$  strip (which is effectively shrinking along $b$). \label{profileICS}}
\end{center}
 \vspace{0pc}
\end{figure}

We also consider the case of electrons arising from WIMP decay
considering a DM signal following linearly the halo profile and with
the same electron injection spectrum as for the $\mu^+\mu^-$
channel. Formally, at the solar position, up to diffusion effects,
exactly the same positron fraction and electron spectrum can be
obtained setting the DM decay rate to $\Gamma=\rho_0\left\langle
\sigma_{\mathrm{A}}v\right\rangle/2m_\chi$. The ICS radiation from
the Halo is however significantly reduced although Fermi can still
discriminate this possibility as shown in Fig.\ref{fig:Skymaps} and
Fig.\ref{profileICS}. At this level, however,the confusion with a
not well understood background could become more problematic
although the peculiar circular shape of the ICS Haze, present also
in this case (see Fig.\ref{fig:Skymaps}), can help to distinguish
the DM signal from the astrophysical background.

Finally, in Fig.\ref{profileICS} we report another forecast
example of the excellent Fermi ability to discriminate among the
astrophysical and annihilating DM scenario considering the
latitude profile and a strip of $60^\circ$ width along $l=0$. We
also show in the upper panel the EGRET data in the same region and
energy range (as derived with the Galplot package (see
also~\cite{Strong:2004de})). Compared with the EGRET data the
annihilation model seems to produce a too much broad peak to fit
the data, beside producing an excessively high normalization. The
decaying model is instead difficult to separate from the
background within the EGRET error bars. With the upcoming Fermi
data at hands, the analysis can be easily generalized to exploit
the full angular shape of the IC Haze. This would clearly offer
the optimal sensitivity to disentangle the different scenarios.

In summary, we have shown that Fermi has the potential to test the
DM interpretation of Pamela/ATIC basically in a model independent
way thanks to the strong IC signal which the Pamela/ATIC electrons
would themselves produce in the galactic halo. The EGRET data seems,
indeed, already to disfavor the DM annihilation interpretation.
Further, the IC signal give rise to a striking ``IC Haze'' feature
peaking around 10-100 GeV which would provide a further mean to
discriminate the DM signal from the astrophysical backgrounds and/or
to check for possible systematics.

\vspace{.3cm} \noindent {\bf Acknowledgements:} G. Miele
acknowledge supports by INFN - I.S. FA51 and by PRIN 2006 ``Fisica
Astroparticellare: Neutrini ed Universo Primordiale" of Italian
MIUR.


\end{document}